\documentclass[pra,twocolumn,amsfonts]{revtex4}
\usepackage{amsfonts}
\usepackage{amsmath}
\usepackage{bm}
\usepackage{epsf}
\usepackage{graphicx}

\def\etal{{\it et al.\/}}

\newcommand{\beq}{\begin{equation}}
\newcommand{\eeq}{\end{equation}}
\newcommand{\beqa}{\begin{eqnarray}}
\newcommand{\eeqa}{\end{eqnarray}}

\newcommand{\la}{\langle}
\newcommand{\ra}{\rangle}

\newcommand{\down}{\downarrow}

\newcommand{\rref}[1]{~(\ref{#1})}

\begin{document}

\title[Ion-vacuum]
{Detection of vacuum entanglement in a linear ion trap }

\author{A. Retzker$^{1}$, J.I. Cirac$^{2}$, and B. Reznik$^{1}$  }

\affiliation{ $^{(1)}$ School of Physics and Astronomy,
Raymond and Beverly Sackler Faculty of Exact Sciences,
Tel-Aviv University, Tel-Aviv 69978, Israel.\\
$^{(2)}$ Max-Planck-Institut
f\"ur Quantenoptik, Hans-Kopfermann-Str. 1, Garching, D-85748, Germany}

\date{\today}

\begin{abstract}
We propose and study a method for detecting ground-state
entanglement in a chain of trapped ions.
We show that the entanglement between single ions
or groups of ions can be swapped to the internal levels of two
ions by sending laser pulses that couple the internal and motional
degrees of freedom. This allows to entangle two ions without
actually performing gate operations. A proof of principle of the effect can be
realized with two trapped ions and is feasible with current
technology.
\end{abstract}

\maketitle


A remarkable phenomenon that appears naturally in quantum field theory,
is that the ground state (vacuum) is entangled, and that 
observables in two separated regions can be entangled. 
Recent studies in quantum information theory have taught us that
entanglement is a physical property which can be exchanged between
systems or used in quantum processes such as quantum
communication, teleportation and quantum cryptography
\cite{review}. This suggests that vacuum entanglement as well
could be detected and used in quantum processes.
There have been several studies of 
vacuum entanglement in field theory \cite{werner,reznik},
as well as in other systems \cite{spin,harmonic,bose}, 
but none have proposed a way to observe vacuum entanglement in a realistic 
experiment. 
The main purpose of this Letter is to suggest a 
realistic physical implementation to observe this phenomenon.


A gedanken-experiment that allows the observation of
vacuum entanglement in field theory has been
suggested \cite{reznik}, which utilizes two basic
ingredients of relativistic field theory and quantum information:
the presence of a  {\em causal structure}, and the
non-increase of entanglement under local operations; operations
performed at two causally disconnected regions do not
increase the entanglement between these regions. Consider two
atoms, $A$ and $B$, which locally interact with the field and with
one another through the long range field interaction. The
interaction with the field can either entangle $A$ and $B$ via the
exchange of propagating quanta, or by transporting vacuum
entanglement into the atoms. Using the fields' built-in causal
structure, one can eliminate the former unwanted process, by
demanding that $cT< L$, where $T$ is the interaction time and $L$
the separation between the atoms. Vacuum entanglement can then
be``swapped" to the atoms' internal levels, which can then be used
for detecting vacuum entanglement.
However this method requires precise
control of the atom-field interaction, which, in the case of an
electromagnetic field, renders the experiment highly
unrealistic. Nevertheless, experimentalists in atomic physics are
taming their systems at the quantum level and can test quantum
effects with the required precision and control for this
type of experiments.
\begin{figure}
 \epsfxsize=3.truein \epsfysize=2truein
\centerline{\epsffile{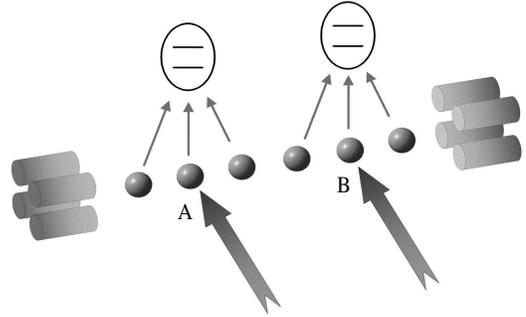}}
  \caption[\, ]
 {Scheme for detecting ground state entanglement:
 the entanglement between two groups of ions is swapped to the
 internal level of ions A and B
 by sending separate laser pulses that induce an interaction between
the internal levels and the position of each ion. }
\label{scheme}
\end{figure}

In the following, we propose and analyze the possibility of observing
vacuum entanglement with trapped ions. We consider a system of
trapped ions that are brought to equilibrium in a linear chain
configuration. The ground state (vacuum) of the system is an
entangled state of the different motional modes (phonons) of the
chain, and manifests entanglement between single ions or distant
groups of ions. In order to detect vacuum entanglement, we
consider processes wherein the external motional degrees of
freedom are mapped to the internal ions states, which are then
used for entanglement detection. The internal levels are well
isolated, they can be temporarily coupled `on demand' to the
phonon modes by sending finite duration laser pulses, and finally
can be measured with nearly perfect precision. In analogy with the
field-theoretical case, the interaction must be limited to a duration
shorter than the time it takes for perturbations to propagate between the two
(probe) ions along the chain. We comment that in the case of ion
chains, this process is interesting on its own, because one can
entangle the internal levels of two ions without actually doing
gates \cite{cirac-zoller}. The most spectacular 
manifestation of the idea would involve a chain with
many ions. However a proof of principle can be attained with just
two trapped ions, and is feasible using current technology.
We shall study both cases and analyze the latter in detail.

We now consider a system of N ions trapped in a linear Paul trap
at very low temperature \cite{book}. The Hamiltonian describing
the ions' motion relative to their equilibrium positions, and internal
levels is  $H_{0}={1\over
2}\omega_z(\sigma_{z}^{(A)}+\sigma_{z}^{(B)}) + \sum \nu_n
a_n^\dagger a_n$. Here $\omega_z$ is the internal levels energy
gap of the two relevant (probe) ions A and B, and $\nu_n$ are the
phonon normal-mode frequencies, with corresponding creation
(annihilation) operators $a^\dagger_n$ ($a_n$). Typically,
$\omega_z$ is in the optical region and $\nu_n\sim \textrm{MHz}$.


We shall begin by analyzing the simplest case with just 
two trapped ions.  The vacuum state is then defined as 
the ground state of the normal modes of the system, i.e.
a product state of the collective and breading modes 
$|0_c\rangle$ and $|0_b\rangle$. In terms of the 
{\em local} single oscillator states
$|n\rangle_{A,B}$, the vacuum is an entangled two mode squeezed
state \cite{modewise}
 \beq |{\rm vac}\rangle= |0_c\rangle|0_b\rangle
= \sqrt{1-e^{-2\beta}} \sum_n e^{-\beta n} |n\rangle_A|n\rangle_B
\label{modewise}
  \eeq
The local number states are the single ion energy eigenstates  
obtained when the displacement of the other ion is set to zero. 
We get
$e^{-\beta}=\sqrt{(\lambda-1/2)/(\lambda+1/2)}$, where
$\lambda=(1/4)[\sqrt{\nu_0/\nu_1}+\sqrt{\nu_1/\nu_0}]$, and
$\nu_0$ and $\nu_1$ are the frequencies of the collective and breathing modes.  Since
$\nu_0/\nu_1=\sqrt3$, we get $\lambda=0.5189$, 
and the von Neumann entanglement \cite{peres} of the squeezed state 
is $E=(\lambda+1/2)\log_2(\lambda+1/2)
-(\lambda-1/2)\log_2(\lambda-1/2)=0.136$ e-bits.

In order to transfer vacuum entanglement into the ion internal
states, we use laser pulses to couple the internal and motional
states of atoms A and B. Close to resonance, with $\omega_{laser}
\approx \omega_z$, the interaction terms for the k'th ion (in the
Lamb-Dicke limit) is given by \cite{review1}
 \beq
   H_{int}^{(k)}=\Omega(t)(e^{-i\phi} \sigma_+^{(k)} +
   e^{i\phi} \sigma_-^{(k)}) x_k \, ,
 \eeq
where  $\sigma_{\pm}$ are the raising and lowering operators,
$\phi$ is the laser phase, and $x_k$ the displacement of the $k$'th
ion. Above we have applied the
rotating wave approximation with respect to the internal levels
but not (as is usually done) to position operators. That is
because, the duration $T$ of the laser pulses satisfies
$1/\omega_z \ll T \le 1/\nu_0$ . The upper bound on $T$ follows
from the requirement that perturbations do not propagate between
the ions during the interaction.

\begin{figure}
   \epsfxsize=2.3truein \epsfysize=1.7truein
\centerline{\epsffile{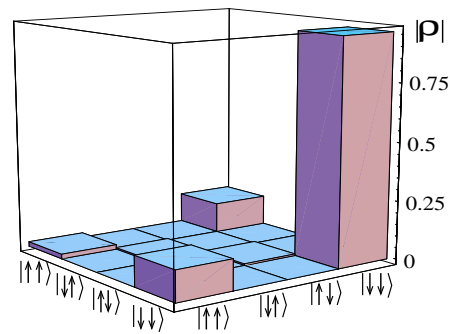}}
\medskip
\caption[\, ]{Histogram of the final density matrix of the
internal levels of ions A and B. The entanglement of formation of
this state accounts for 97\% of the computed initial ground state
motional entanglement. } \label{swap}
 \end{figure}

Using the available interaction
(2), we would like to swap the motional entanglement into
the internal ion levels which are initially prepared in a non-entangled state
$|\downarrow\rangle_A |\downarrow\rangle_B$.
The smallness of $e^{-\beta}$ implies that the entanglement
arises mostly from the first two terms of \rref{modewise}. 
We therefore seek a procedure, of typical duration  
$T\ll 1/\nu_0$, that maps  
\begin{equation}\label{final}
|\rm vac\rangle |\downarrow\rangle |\downarrow\rangle \to
|\chi\rangle[ |\downarrow\rangle |\downarrow\rangle + e^{-\beta}
|\uparrow\rangle |\uparrow\rangle]
\end{equation}
where $\chi$ is the final state of the ions. 
The interaction then acts separately on each ion, and 
swaps the lowest two motional states $|0\ra$ and 
$|1\ra$ to the two ion internal states. In this $4\times 4$ subspace the
map  is expected to approximate the unitary swap transformation 
$e^{i \pi/4 (\tilde\sigma_x \sigma_x + \tilde\sigma_y \sigma_y)}$, 
where $\tilde\sigma_x=|0\ra\la 1|+|1\ra\la0|$ and $\tilde
\sigma_y=i|1\ra\la 0|-i|0\ra\la1|$ act on the number states.
(We have ignored a $\tilde\sigma_z\sigma_z$ term 
which in our case adds a trivial phase since the initial state is a 
$\vert\down\rangle \vert\down\rangle$ state).
We note that $\tilde\sigma_x$ and $\tilde\sigma_y$
can be approximated by $x$  and $p$ respectively. We therefore expect that
the swap will require coupling of the internal levels with 
both position and momentum, since the truncation of the $x$ operator yields
the $\tilde\sigma_x$ operator and truncation of momentum yields the 
$\tilde\sigma_y$ operator.
Based on this intuition we proceed with the following construction. 
We consider the following sequence of unitary
operations, $U_s= V(\alpha_1)W(\beta_1)V(\alpha_2)W(\beta_2)\cdots
V(\alpha_n)W(\beta_n)$, where
 \beq V(\alpha) = e^{i\alpha \sigma_x
x} \, \ \ \ \ \ \ W(\beta)=e^{i\beta \sigma_y p} \, ,
 \eeq to be
performed on each ion separately by  sending a sequence of laser
pulses. The $V(\alpha)$ evolution, can be obtained by sending a
laser pulse of duration $T$ and phase $\phi=0$, such that $T\ll
1/\nu_0$ and $\int \Omega(t)dt =\alpha$. In order to generate a
$W(\beta)$ evolution, we set the laser phase to $\phi=\pi/2$, and
allow the system evolve freely for short time interval $dt=\tau$
in between a pair of pulses. Denoting $V'(\beta) = \exp(i\beta
\sigma_y x)$, we obtain:
   \beqa V'_{t=\tau}(-\beta')
V'_{t=0}(\beta')&=& e^{-i(\beta' \sigma_x (x + {p\tau\over m})+
O(\nu^2\tau^2)}
e^{i\beta' \sigma_x x} \nonumber \\
&=&e^{-i{\beta'\over m}(\sigma_x p\tau+ {1\over 2}\tau\beta')}+O(\nu^2\tau^2)
  \eeqa
where we have used the approximation $x(\tau) = x(0) + p(0)\tau/m
+ O(\nu^2\tau^2)$. (Alternatively, in the Schrodinger picture we
notice that $V^\dagger e^{-iH_{phonon}t}V $ shifts the kinetic
term as $p^2 \to (p+ \beta' \sigma_y)^2 = p^2 + 2\beta' \sigma_y p
+ \beta'^2$.) Taking the limit  $\nu^2\tau^2  \ll 1$, and
maintaining $\beta\equiv \beta'\tau/m = O(1)$ we obtain the
required effective coupling to $p$. Therefore the sequence of $n$
pairs of $V(\alpha)W(\beta)$ pulses, can be generated by $3n$
ordinary pulses with $n$ free evolution intermediate intervals of
total duration $dt=n\tau$. By optimizing the entanglement of
formation \cite{EOF}, $E_F(\alpha_i,\beta_i)$, over the free
parameters $\alpha_i$ and $\beta_i$ the
transformation (\ref{final}),
where $\chi$ is the final motional state of the ions, can be
generated with high efficiency. This transformation swaps the
first two terms in Eq. (\ref{modewise}) to the ion's internal level states.
After a sequence of three $VW$ pulses, the entanglement of
formation
 of the final internal level state contains 97\% of the
computed ground-state entanglement. The optimal sequence is in
this case $V(0.31)W(0.38)V(0.50)W(0.39)V(0.53)W(0.16)$. (With two
pulses we get at most 93\%).  Expressed in 
the relevant $4\times 4$ subspace, this unitary operation has indeed a 
a structure which closely resembles a swap. 
Testing the purity of the final state
$\rho_{AB}$, we find  ${ tr} \rho_{AB}^2 =0.997$. The final
density matrix of the internal levels is depicted in Fig. \ref{swap}. The
measurement precision for the density matrix in recent experiments
is about $1\%$ \cite{accuracy}, and hence sufficient for observing
the entanglement of the final state.  

We comment that an alternative to the above approach could be to
separate the two ions by moving them apart. This effectively
"turns off" the interaction between them, allowing for
longer duration of the detection process. It is then easier to
generate the desired swap using continuous on-resonance laser
pulses \cite{kraus}. In order not to affect the entanglement
between the ions, the separation has to be fast compared with the
propagation time scale $~1/\nu_0$. This can be done by increasing
the potential between the ions. The
possibility of changing the local potential and moving ions,
without effecting the internal ion states, has been recently
demonstrated experimentally
 \cite{wineland2004,rowe2001}.

\begin{figure}
   \epsfxsize=2.0truein \epsfysize=1.4truein
\centerline{\epsffile{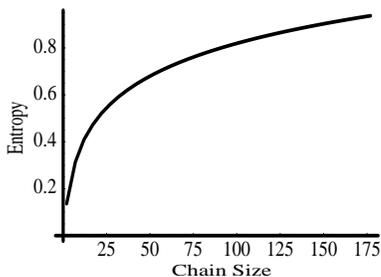}}
\caption[\, ]{Entanglement (in e-bits) between complementary
symmetric groups of ions as a function of the total ion number. }
\label{entropy}
 \end{figure}

We next turn to the more general case of entanglement in  
a chain with $N$ ions, and consider first the entanglement properties
of the system.
Since the vacuum is a pure Gaussian state, the Schmidt decomposition 
for two complementary parts of the chain can 
be expressed as a direct product of squeezed states \cite{modewise}.  
The von-Neumann bi-partite entanglement can then be obtained 
by summing the entanglement contribution of each squeezed state 
(Fig. \ref{entropy}).

\begin{figure}
   \epsfxsize=2.0truein \epsfysize=1.4truein
\centerline{\epsffile{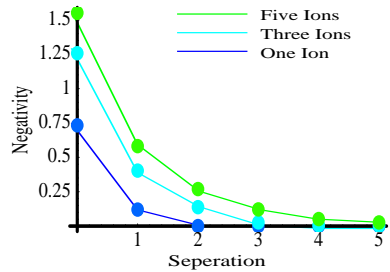}}
\caption[\, ]{Logarithmic Negativity  between two groups consisting of 1,3, and
5 ions, as a function of their separation in a chain of 20 ions.}
\label{ent}
 \end{figure}

The state of two {\em sub-groups}, $\tilde A$ and $\tilde B$, each
consisting of $n_A$ and $n_B$ ions,  separated by $l_s$
ions, is described by a reduced Gaussian density
matrix $\rho_{\tilde A\tilde B}={tr}_{l\notin \tilde A,\tilde B}
(|{\rm vac}\rangle\langle {\rm vac}|)$. The entanglement between the groups 
(Fig. \ref{ent}) can be characterized by the Negativity \cite{negativity,negativity1}.
It vanishes for separation larger then one. However as the group
size increases, it persists for larger separations.


\begin{figure}
  \centerline{ \epsfxsize=1.6truein \epsfysize=1.8truein
\epsffile{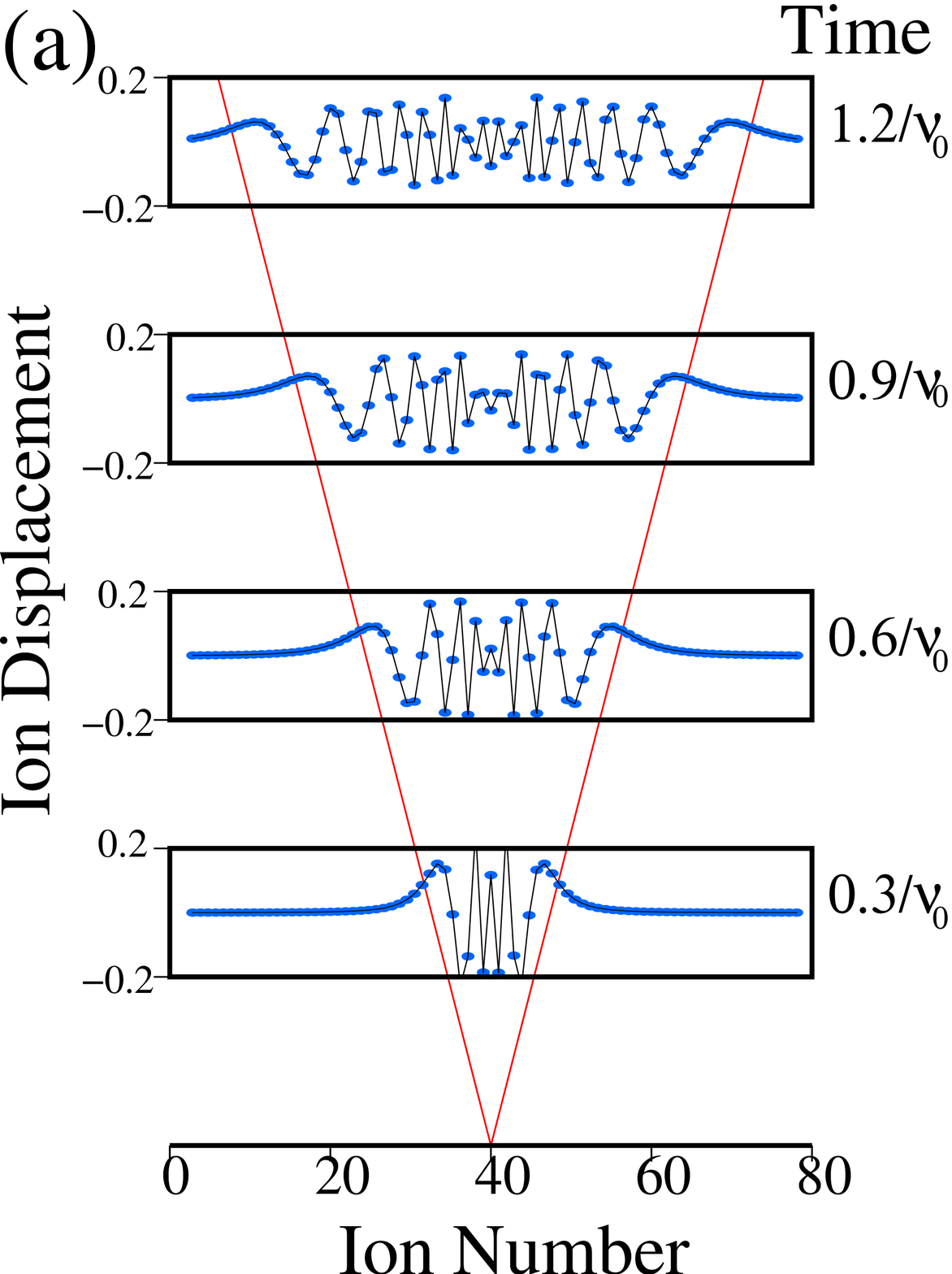} \ \ \ \  \epsfxsize=1.6truein
 \epsfysize=1.8truein\epsffile{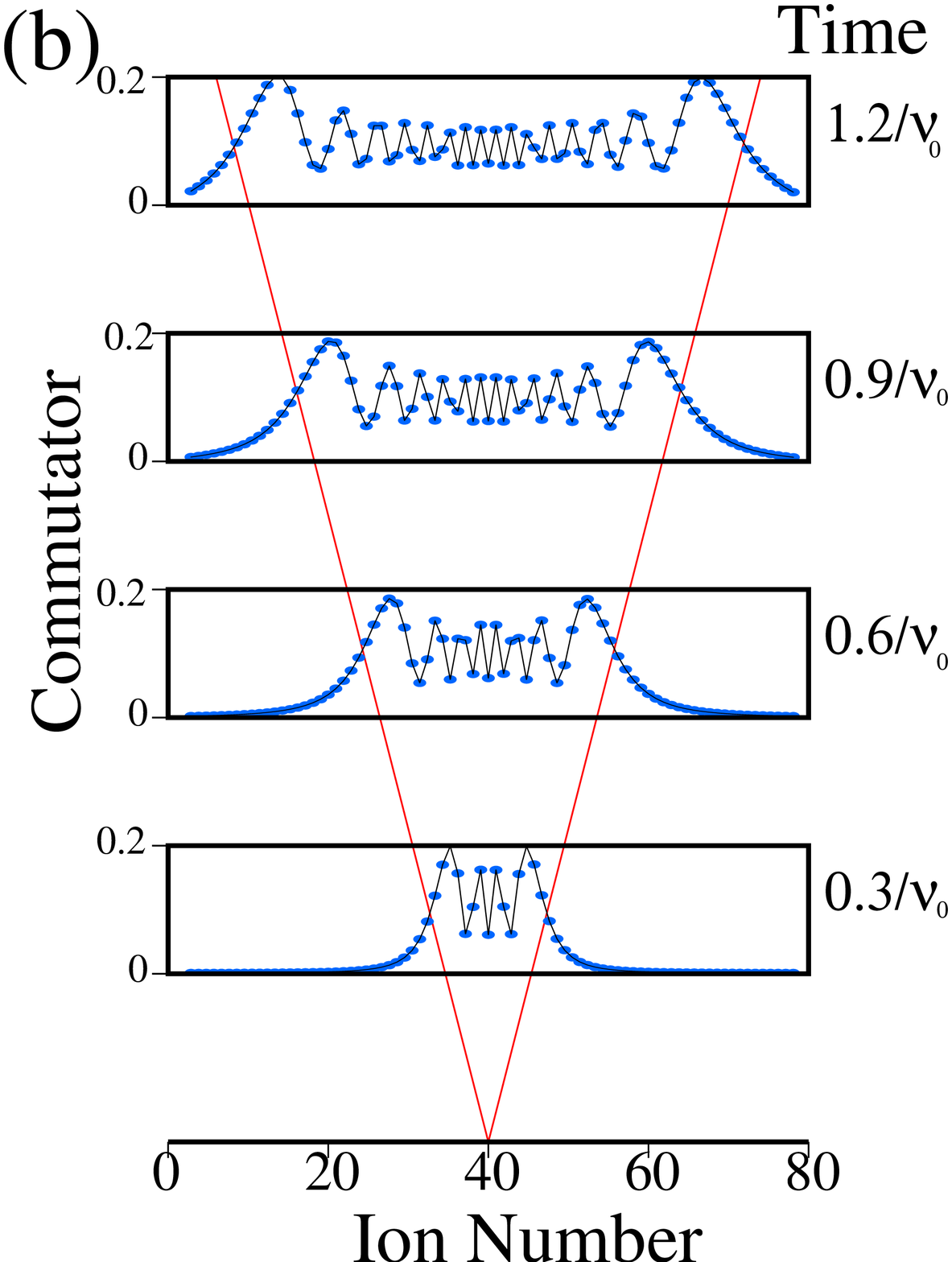} }
\medskip
\caption[\, ]{  a) Classical propagation of a perturbation
originating at the center of the chain.  b) The commutation
relation between the displacement operators of the $n$'th ion and
the central ion in a chain of 80 trapped ions, at different time
slices.}\label{prop}
 \end{figure}

We shall examine the possibility of detecting the entanglement
by coupling to the internal levels of two ions, A and B, for 
a time duration $T$.
We can check that as long as $T<1/\nu_0$, 
the interaction with respect to two complementary 
parts of the chain can be regarded as local. This is seen by noticing
that the evolution operator, $U(T)$, can be factorized in
the interaction picture as
   \beq U = U_A \otimes U_B \otimes
   e^{-i/2\int dtdt' f(t,t')\sigma_{AB} } \, ,
   \eeq
where $U_{k}$ act on A or B, $\sigma_{AB} \equiv
\Pi_{k}(e^{-i\phi} \sigma_+^{(k)} + e^{i\phi} \sigma_-^{(k)})$,
and $f(t-t') = [x_A(t),x_B(t')]$. The last term above, involving
$\sigma_{AB}$, is a unitary operator that can increase
entanglement ``non-locally". However, as can be seen in Fig. \ref{prop},
the non-commutativity described by $f(t-t')$ vanishes rapidly, and
for sufficiently short interaction time, or large enough spatial
separation, this non-causal effect is suppressed.

We begin with the initial ground-state
$|{\rm vac}\rangle|\downarrow\rangle|\downarrow\rangle$,
and proceed to evaluate the reduced state  $\rho_{AB}(T)=
tr[U(T)|I\rangle\langle I| U^\dagger(T)]$ perturbatively.
Assuming that the intensity of the laser pulses is sufficiently weak,
we expand $U(T)$ in a power series, and to lowest
order in  $\Omega$ we obtain
\begin{equation}
\rho_{AB}=
\left(
\begin{array}{cccc}
 \scriptstyle\Vert X_{AB}\Vert ^2       &      0         &      0       &
\scriptstyle - \la 0| X_{AB}\ra  \\
      0             &\scriptstyle \Vert E_{A}\Vert ^2    &\scriptstyle\la E_B|E_A\ra&
0\\
      0             &\scriptstyle\la E_A|E_B\ra  &  \scriptstyle\Vert E_B\Vert ^2   &
0 \\
 \scriptstyle - \la X_{AB}|0\ra &      0         &      0   & 1\scriptstyle - \Vert
E_{A}\Vert ^2-\Vert E_B\Vert ^2
\end{array} \right)\scriptstyle
\label{density}
\end{equation}
Here $|E_A\ra= X_A|\rm vac\ra$, $|X_{AB}\ra= X_A X_B|\rm vac\ra$
and $X_k = \int dt\Omega(t)e^{ i\delta t}x_k(t)$, ($k=A,B$), and
$\delta$ is the detuning. Using the Peres-Horodecki separability
criterion, it then follows that $\rho_{AB}(T)$ is entangled iff
$\mathcal{N}(\rho_{AB})\approx|\la 0|X_{AB}\ra| - \Vert E_A\Vert
\Vert E_B\Vert >0$, where $\mathcal{N}(\rho_{AB})$ is the
negativity. This condition can be understood physically as the
requirement that the virtual off-shell single phonon exchange
process (described by $X_{AB}$), is sufficiently large to overcome
the decoherence effects due to local phonon emission (described by
the $E_{A,B}$ terms).

\begin{figure}
  \centerline{ \epsfxsize=1.5truein \epsfysize=1.5truein
\epsffile{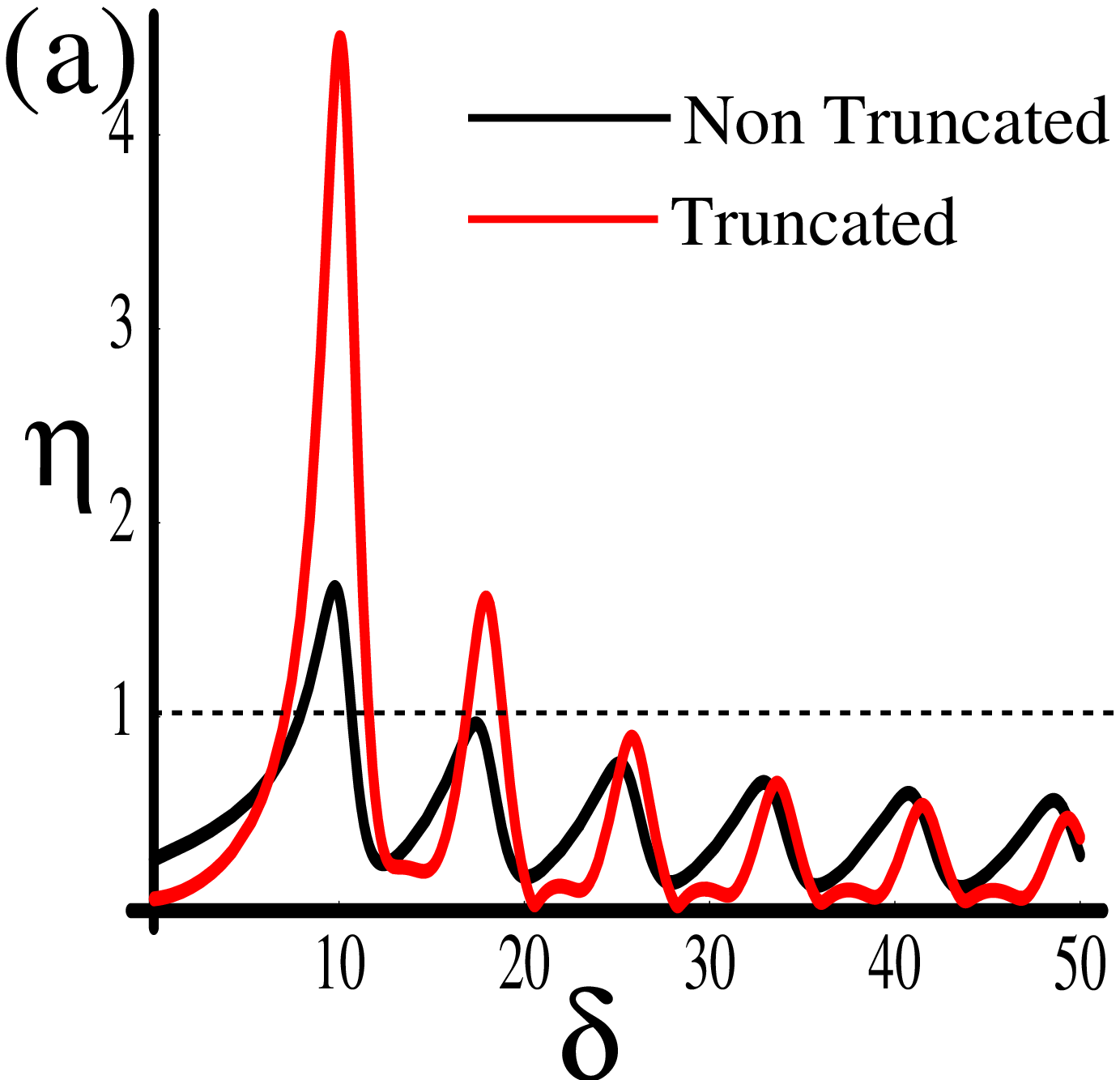} \ \ \ \  \epsfxsize=1.7truein
 \epsfysize=1.5truein\epsffile{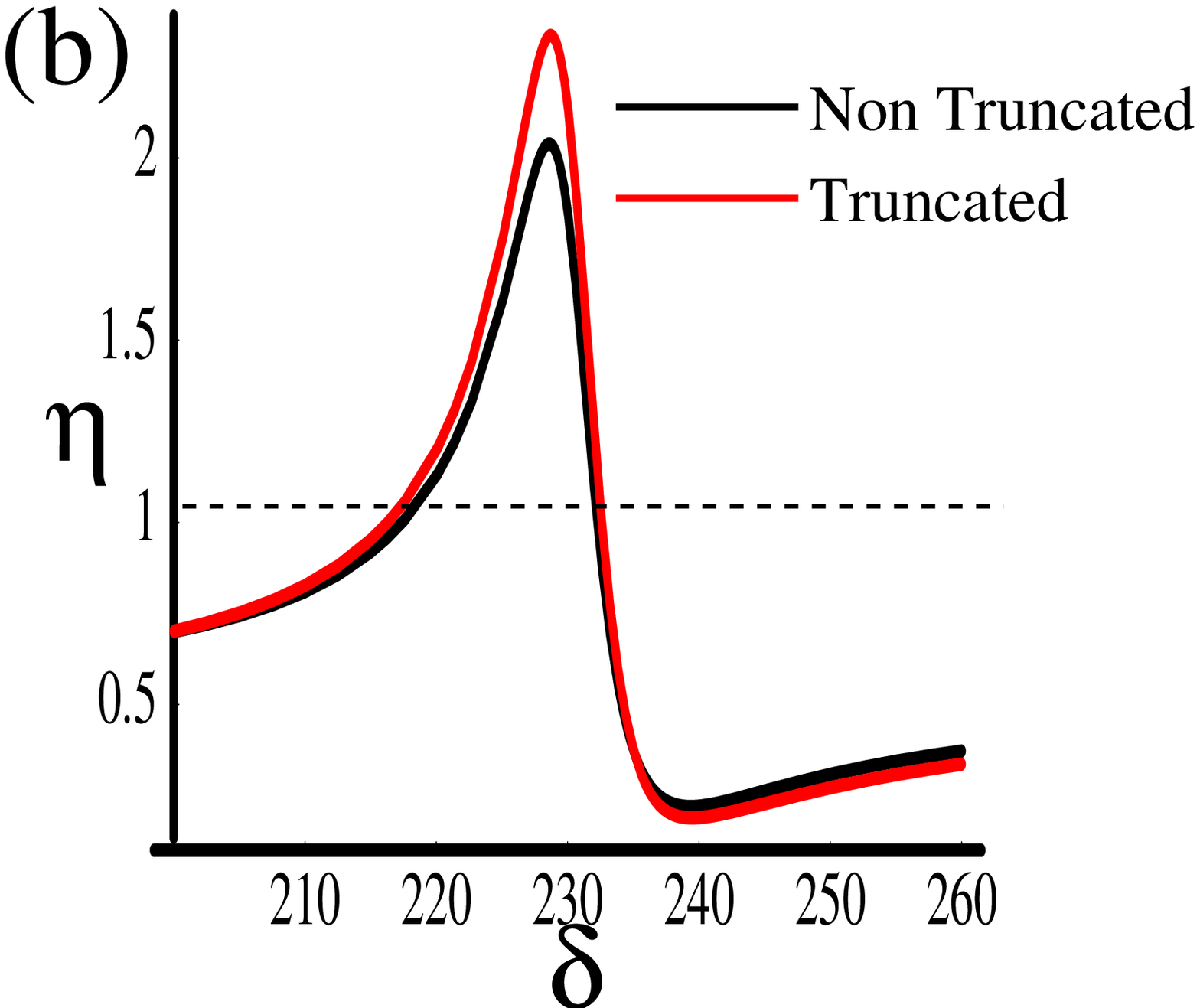} }
\medskip
\caption[\, ]{The ratio $\eta$ for ions in a chain of $N=20$ ions
as a function of the detuning $\delta$. a) The ion probes are
located at $l=6,15$ and $T=0.8$ (in unites wherein $\nu_0=1$). b)
$l=10,11$ and $T=0.05$. The range $\eta>1$ signifies entanglement.
} \label{ent1}
 \end{figure}

To verify that the above condition amounts to a detection of vacuum
entanglement, rather than a direct interaction due to the
non-local correction in Eq. (3), we have repeated the computation,
using the same initial state $|I\ra$, but with a modified
truncated evolution. In the latter truncated case, we
``disconnected" the chain by eliminating the interaction between
ions at the different halves, $n>N/2$ and $n<N/2$, of the chain.
This can be easily achieved by replacing the potential term in the
free phonon Hamiltonian by  $\sum x_i G_{ij} x_k \to \sum x_i
G_{ij}^T x_k$ where $G^T= G_A\oplus G_B$ is block-diagonal. This
truncated evolution, does not change the entanglement between the
two halves of the chain since an exact separability holds in Eq.
(3),
 (i.e. $f(t-t')=0$). The ratio $\eta=|\la 0|X_{AB}\ra|/ \Vert
E_A\Vert \Vert E_B\Vert$  is plotted Fig. \ref{ent1} as a function of the
detuning $\delta$, for $N=20$ ions. In Fig. \ref{ent1}a the probe-ions are
situated at sites $l_A=6$ and $l_B=15$, and we can see that A and
B become entangled ($\eta>1$) in a range of frequencies. Since
small violations of causality are expected in the non-truncated
model, one could have anticipated that the non-truncated case
should give rise to more entanglement. On the contrary, we see
that it is the truncated case which gives rise to more
entanglement. To understand this consider first Fig. \ref{ent1}b, in which
nearest neighbor
 ions $l_{A,B}=10,11$ have been used as probes.
We find that the truncated and non-truncated models precisely
agree for sufficiently small $T\ll 1/\nu_0$. In this case, since
there is  pre-existing local entanglement between the close ions,
the probe can detect entanglement in an arbitrarily short time, and
the truncated interaction has here no effect because evolution is
not required. On the other hand, in the case of separated ions,
propagation effects ``communicate" between the probes and ions
closer to the center of the trap which carry the most entanglement.
This suggests that the larger entanglement in the truncated model
is due to perfect wave reflection at the boundary between the
regions.

In conclusion, we have proposed an efficient method for detecting
vacuum entanglement by mapping motional states of trapped
ions or groups of ions to the ions' internal levels. 
It is remarkable that this phenomenon, which can be considered 
as purely fundamental, may also help to realize quantum information tasks.
Further investigations will determine whether 
vacuum entanglement can be used to entangle
internal degrees of freedom in a fast way, 
and to produce spin squeezing, which
would be of practical interest in, for example,
atomic clocks \cite{wineland94}.



 AR and BR acknowledge the support by ISF, grant
62/01-1. JIC was supported by EU projects, the DFG, and the
Bayerischen Staatsregierung.


\begin{thebibliography}{99}
\bibitem{negativity1} The logarithmic negativity measures the degree of which the partial transposed 
density matrix (which is a transpose of the density matrix with respect to one of the  subsystems)
fails to be positive, moreover it bounds the distillable entanglement.
\end{thebibliography}
\end{document}